# A Weak Fermi Gamma-ray Event Associated with a Halo CME and a Type II Radio Burst

Nat Gopalswamy*[(1)], Pertti Mäkelä[(2)], and Seiji Yashiro[(2)]
(1) NASA Goddard Space Flight Center, Greenbelt, MD 20771, USA, https://cdaw.gsfc.nasa.gov
(2) The Catholic University of America, Washington DC 20064, USA

## Abstract

We report on the 2015 June 25 sustained gamma-ray emission (SGRE) event associated with a halo coronal mass ejection and a type II radio burst in the decameter-hectometric (DH) wavelengths. The duration and ending frequency of the type II burst are linearly related to the SGRE duration as found in previous works involving intense gamma-ray events. This study confirms that the SGRE event is due to protons accelerated in the shock that produced the DH type II burst.

## 1 Introduction

The origin of energetic protons responsible for neutral-pion gamma-ray continuum from the Sun observed beyond the flare impulsive phase has been an unsolved problem for more than 30 years [1]. Fermi's Large Area Telescope (Fermi/LAT [2]) data revealed that such events at energies >100 MeV are rather common [3-5]. In some cases, the gamma-ray duration can be almost a day [6-8]. Since the flare is long gone before the gamma-ray emission ends, these events are now known as sustained gamma-ray emission (SGRE) events [7,9] or late phase gamma-ray emission (LPGRE) [5]; the traditional name has been long duration gamma-ray flare (LDGRF) [10]. The SGRE event results from the precipitation ≥ 300 MeV protons to the solar atmosphere resulting in neutral pions that promptly decay into gamma-rays. The energization mechanism of these protons has been controversial: flare reconnection or shock acceleration. Under the flare reconnection paradigm, the long duration of the gamma-ray events is explained by protons trapped in flare loops slowly precipitating to the solar atmosphere [11]. In the shock mechanism, the shock continuously accelerates protons that diffuse back to the solar surface to produce the gamma-rays [12]. Particles escaping from the shock into space are detected as solar energetic particle (SEP) events. It has been difficult to decide between these two competing mechanisms for lack of distinguishing characteristics.

SGRE events observed by Fermi/LAT have been found to be associated with fast CMEs and DH type II bursts [5]. Both these phenomena are closely related to fast mode shocks, so there is some basis to consider the shock mechanism. The recent result that linked the SGRE duration to the DH type II burst duration turned out to be the crucial missing link for the shock mechanism [8]. If flare particles trapped in flare structures are responsible for the late phase emission, there should be no correlation between the durations of type II bursts and SGRE. Thus, while the impulsive phase gamma-rays are certainly due to protons accelerated in the flare site, the late phase emission has to be due to the shock. To firm up this result, we are currently investigating why there are a lot more DH type II bursts than the number of SGRE events. During this investigation, we came across a weak SGRE event that occurred on 2015 June 25. In this paper we show that even this weak event has the same quantitative relation with the associated DH type II burst and hence fully consistent with the shock paradigm.

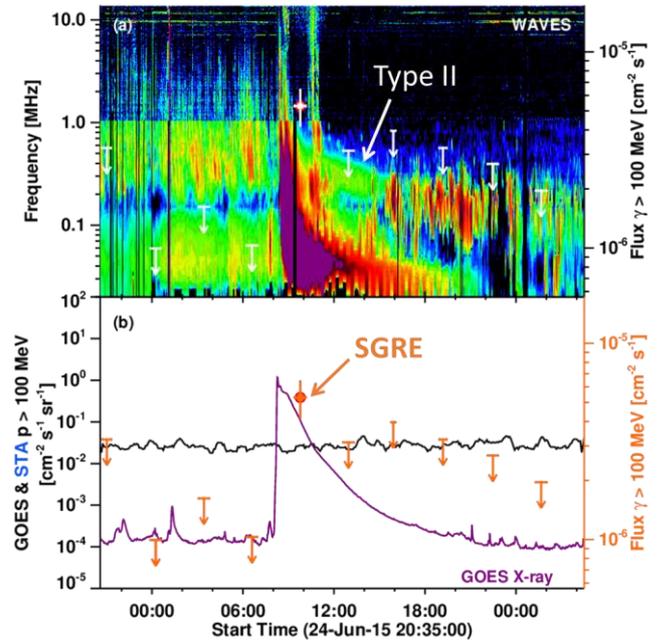

**Figure 1.** Wind/WAVES dynamic spectrum (top) showing the type II radio burst with superposed SGRE flux at energies >100 MeV obtained from the Maximum Likelihood Method. The GOES X-ray light curve and >100 MeV proton intensity along with the >100 MeV gamma-ray flux (bottom). If there is STEREO-A (STA) proton data it is plotted, but no data for this event. The SGRE has one significant data point (marked by arrow); other points denote the background level (~$1.66\times10^{-5}$ cm$^{-2}$ s$^{-1}$). The peak SGRE flux is only $2.21\times10^{-5}$ cm$^{-2}$ s$^{-1}$ computed using the light bucket method [5].

## 2 The SGRE Event

The SGRE was observed by Fermi/LAT in association with a M 7.9 soft X-ray flare that started, peaked, and ended at 08:02 UT, 08:16 UT, and 09:05 UT on 2015 June 25. As in all >3-hr SGRE events reported so far [13], this SGRE was also associated with a halo coronal mass ejection (CME) observed by the Large Angle and Spectrometric Coronagraph (LASCO, [13]) on the Solar and Heliospheric Observatory (SOHO). The halo CME first appeared in the LASCO field of view at 08:36 UT. The Radio and Plasma Wave (WAVES) experiment [14] on board Wind detected a type II radio burst in association with the SGRE event. Details on the type II burst are available in the Wind/WAVES catalog of type II bursts: https://cdaw.gsfc.nasa.gov/CME_list/radio/waves_type2.html [15]. The type II burst appeared at the highest frequency (~14 MHz) of the WAVES spectral band at 08:35 UT and continued below 200 kHz over the next several hours. The type II burst also had a metric component starting at 08:16 UT. The SGRE event was also associated with an SEP event also cataloged at the CDAW Data Center (https://cdaw.gsfc.nasa.gov/CME_list/sepe/).

## 2.1 The Duration Comparison

The close relationship of SGRE durations with type II burst durations and ending frequencies has been reported for SGRE events with duration >3 hr [16]. The correlations indicate that for longer-duration SGRE events, the underlying shock travels a larger distance from the Sun, accelerating protons and electrons. It is natural to ask how the current weak event would fit in this relationship. The SGRE duration is measured from the peak of the associated soft X-ray flare (08:16 UT) to a data point after the last signal data point [8]. In the present event, there is only one signal data point (09:45 UT), so the duration is determined by this data point and the next data point at 13:00 UT. The time elapsed from 08:16 UT to the midpoint (11:22:15 between 09:45 UT and 13:00 UT as 3.1±0.79 hr. The DH type II burst starts around 08:35 UT and ends somewhere between 14:00 and 15:25 UT, so the duration is 6.13±1.38 hr. The other parameter we need is the ending frequency of the type II burst, which is ~250±100 kHz (see Fig. 1).

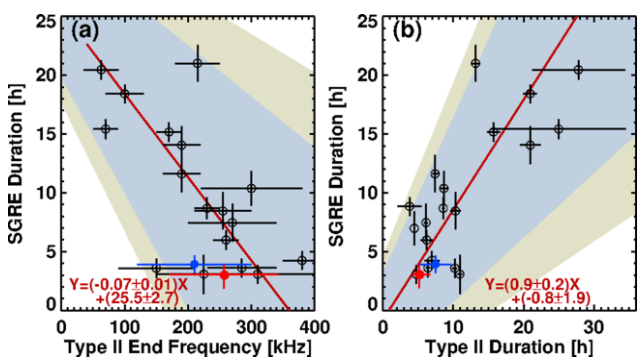

**Figure 2.** (a) Scatter plot between SGRE duration and type II ending frequency for 19 events with duration >3 h. The blue data point is the backside event on 2014 September 1 whose data points are not included in the correlation. The red data point corresponds to the 2015 June 25 SGRE event. (b) Scatter plot between SGRE duration and type II duration with the blue and red data points having the same meaning as in (a). The shaded areas correspond to 95% and 99% confidence intervals. The 2015 June 25 event lies within the 95% confidence interval. The linear fits to the data points are shown on the plots.

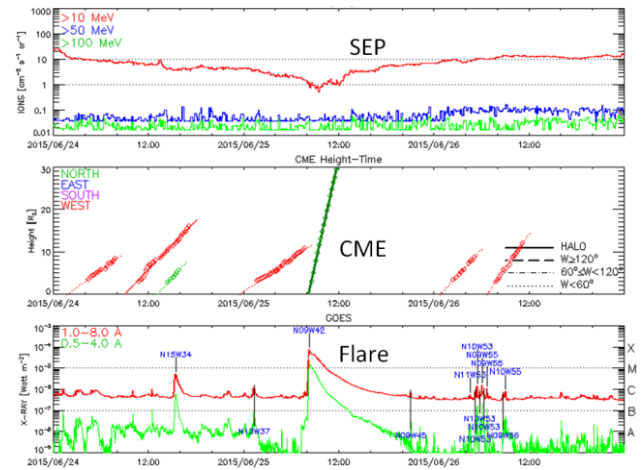

**Figure 3.** SEP intensity in the three GOES energy channels >10 MeV, >50 MeV and >100 MeV (top), the CME height-time plot (middle) and the GOES soft X-ray intensity in the 1-8 Å and 0.5-4 Å channels showing the flare (bottom).

## 2.2 The SEP Event and CME Trajectory

One of the key requirements of an SGRE event is the acceleration of protons to energies >300 MeV. The >100 MeV proton intensity in Fig. 1 is flat and at insignificant levels. The SEP event was large with a peak >10 MeV intensity of ~22 pfu. Figure 3 shows the SEP event in the integral channels >10 MeV, >50 MeV, and >100 MeV. The >10 MeV intensity increases from an elevated level due to previous events and crosses the 10 pfu level early on 2015 June 26. There is also a tiny increase in the >50 MeV channel, but almost nothing in the >100 MeV channel indicating a soft spectrum. For the production of SGRE, one needs >300 MeV protons and we typically use the >100 MeV intensity as proxy for them. Even though supposedly well-connected to the SEP source (N09W42), the GOES satellite did not detect >100 MeV protons. So, what is the source of the >300 MeV protons?

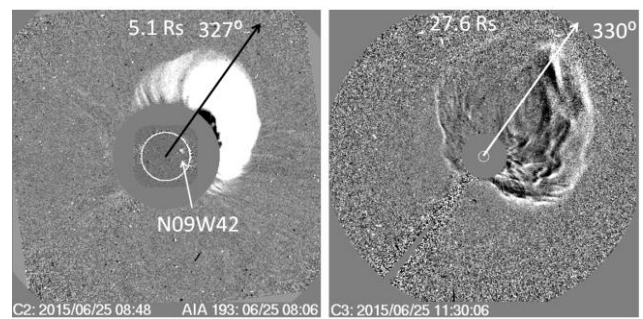

**Figure 4.** (left) SOHO/LASCO/C2 difference image at 08:48 UT showing the CME heading along position angle (PA) 327º (black arrow). The EUV 193 Å difference image from the Atmospheric Imaging Assembly (AIA) [17] on Solar Dynamic Observatory (SDO) shows the bright flare structure at N09W42. The CME leading edge at this time

is at ~5.1 Rs. For a source at N09 latitude, the CME is expected to be heading radially out close to the equator (PA~270º). (right) LASCO/C3 difference image showing the leading edge along PA=330º at a height of 27.6 Rs at 11:30 UT (close to the end of the SGRE event).

Poor correlation between SEP events and SGRE events is known before [8]. The best examples are the 2011 March 7 and 2012 January 23 SGRE events. These events also did not have significant levels of >100 MeV protons. It was found that the noses of these CMEs were at high latitudes and hence not well connected to an Earth observer. The source of high-energy particles is confined to the nose region of the shocks, so they are not detected at Earth, but they do precipitate at the Sun and produce gamma-rays. Figure 4 shows that the nose of the 2015 June 25 CME is at PA~330º, which is about 60º from the ecliptic and hence not connected to Earth. The coronagraph observations thus confirm that the poor latitudinal connectivity is responsible for the lack of high-energy particles at Earth even though they likely are accelerated at the nose region and precipitate to the atmosphere and produce the SGRE.

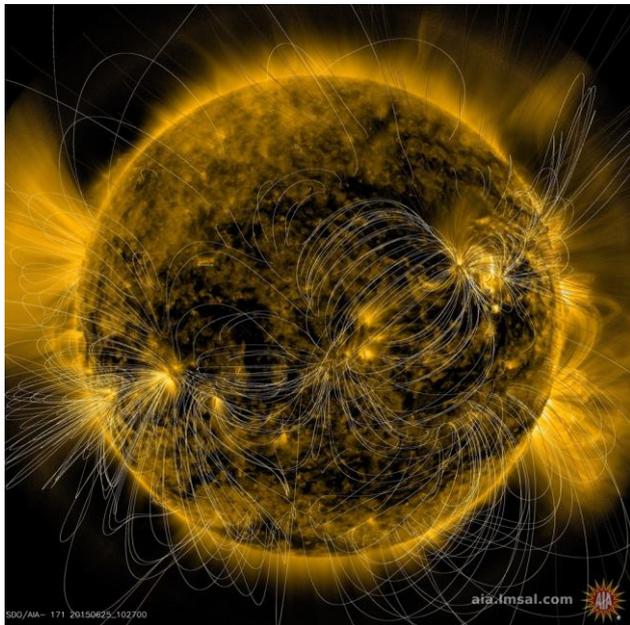

**Figure 5**. SDO/AIA 171 Å image showing the post-eruption arcade in the northwest quadrant of the Sun. Superposed on the image are magnetic field lines obtained from potential field source-surface extrapolation. The large-scale closed field structure to the east of the eruption region is due to field lines emanating from AR 12371 and connecting to an extended negative polarity region close to the disk center. There are also coronal holes to the northwest and southeast of the eruption region.

Figure 5 shows the magnetic environment of the eruption region (in SDO/AIA 171 Å image) indicated by the magnetic field lines derived from the potential field extrapolation of the photospheric field. We see that the eruption region is surrounded by complex field structures, except for the northwest side that corresponds to the trajectory of the CME. CMEs are known to be deflected by open and closed magnetic structures representing large magnetic pressure gradients [18]. Figure 5 provides a qualitative explanation for the non-radial propagation of the CME causing the poor latitudinal connectivity.

The 2015 June 25 SGRE event was associated with fast halo CME shown in Fig. 4 (right). The sky-plane speed was 1627 km/s, which becomes 1805 km/s when deprojected (https://cdaw.gsfc.nasa.gov/CME_list/halo/halo.html) using a cone model. The 3D speed is very close to the average value of >3-hr SGRE events (~2000 km/s). It is remarkable that all the SGRE events, including the current one, are associated with halo CMEs, which are more energetic on the average. The SGRE ends when the CME/shock is still within the LASCO field of view. In intense SGRE events, the shock is located halfway between the Sun and Earth when the SGRE ends.

## 3 Discussion

We investigated the CME, type II radio burst, and SEP event associated with the 2015 June 25 SGRE event. The event was identified during our reverse investigation to find out the lack of SGRE events in many DH type II burst events. In fact, we have identified several SGRE events simply based on the type II bursts observed in the DH wavelength domain. It is remarkable that this extremely weak event has properties and associated phenomena similar to that of the larger events. The SGRE fluence of this event is only ~0.05 cm$^{-2}$.

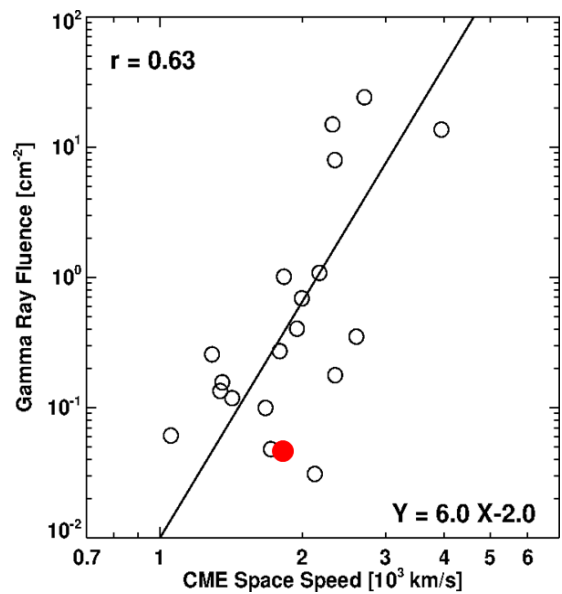

**Figure 6**. Scatter plot between the CME 3D speed and the SGRE fluence for a set of 19 events reported in [16]. The red data point corresponds to the 2015 June 25 SGRE event studied in the present paper. In the regression line shown on the plot, Y is the SGRE fluence and X is the logarithm of the speed in units of 1000 km/s.

Figure 6 shows the scatter plot between SGRE fluence and CME speed for a set of 19 SGRE events reported in [16]. The fluence of the 2015 June 25 event is shown on the scatter plot paired with the CME speed (1800 km/s). If we substitute X=log (1.8) in the regression line, we get a fluence of ~0.34 cm$^{-2}$, which is a factor of ~7 larger than observed. This value is still within the scatter seen in Fig. 6, but the deviation is likely to be due to the uncertainty in the 3D speed obtained by a simple cone-model deprojection.

It is interesting to note that the current CME is one of a series of 5 halos observed from the same active region (AR 12371) between 2015 June 18 and July 1 as the region rotated from E47 to W118 (see Table 1). Of these, the halo on 2015 June 21 was associated with an SGRE event that had the highest fluence among the >3 hr SGRE events [16]. A closer examination reveals that the type II burst was of broad band and intense only in the two halos that resulted in the SGRE (see Fig. 1 and [16]). Furthermore, the 3D speed ($V_{CME}$) is the highest for the 2015 June 21 and 25 events. While the simple deprojection resulted in $V_{CME}$= 1740 km/s, a flux rope fit resulted in a shock speed exceeding 2000 km/s [19]. Thus, the nature of the type II burst and CME kinematics are consistent with the occurrence of SGRE. The slower halos with weaker type II bursts were not associated with an SGRE event.

Table 1. The five halo CMEs from AR 12371

| Date | Location | Flare | $V_{CME}$ | SGRE? |
|---|---|---|---|---|
| June 18 | N12E47 | C3.5 | 1398 | No |
| June 21 | N12E13 | M2.6 | 1740 | Yes |
| June 22 | N12W08 | M6.5 | 1573 | No |
| June 25 | N09W42 | M7.9 | 1805 | Yes |
| July 01 | N09W118 | ???? | 1435 | No |

In summary, the 2015 June 25 SGRE event had all the typical signatures of an SGRE event: a fast halo CME, a metric to kilometric type II radio burst, and a large SEP event. Therefore, the shock mechanism seems to operate even in the weak events. It is significant that the SGRE event was identified based on the existence of type II burst, further strengthening the shock connection. The SGRE event is a strong evidence for the presence of >300 MeV protons in the event. The SEP event observed at Earth was of soft spectrum with not many high-energy particles observed because of the non-radial propagation of the associated CME.

# 6 Acknowledgements

We benefited from the open data policy of *SOHO*, *STEREO*, SDO, *GOES*, and *Wind* teams. Work supported by NASA's LWS and Heliophysics GI programs.